\documentclass[conference]{IEEEtran}
\usepackage{amsmath}
\usepackage{graphicx}
\usepackage{caption} 
\usepackage{subcaption}
\usepackage{tikz}

\usepackage{background}
\usepackage{blindtext}
\backgroundsetup{scale = 1, angle = 0, opacity = 0.2,
   contents = {\includegraphics[width = \paperwidth,
   height = \paperheight, keepaspectratio]
   {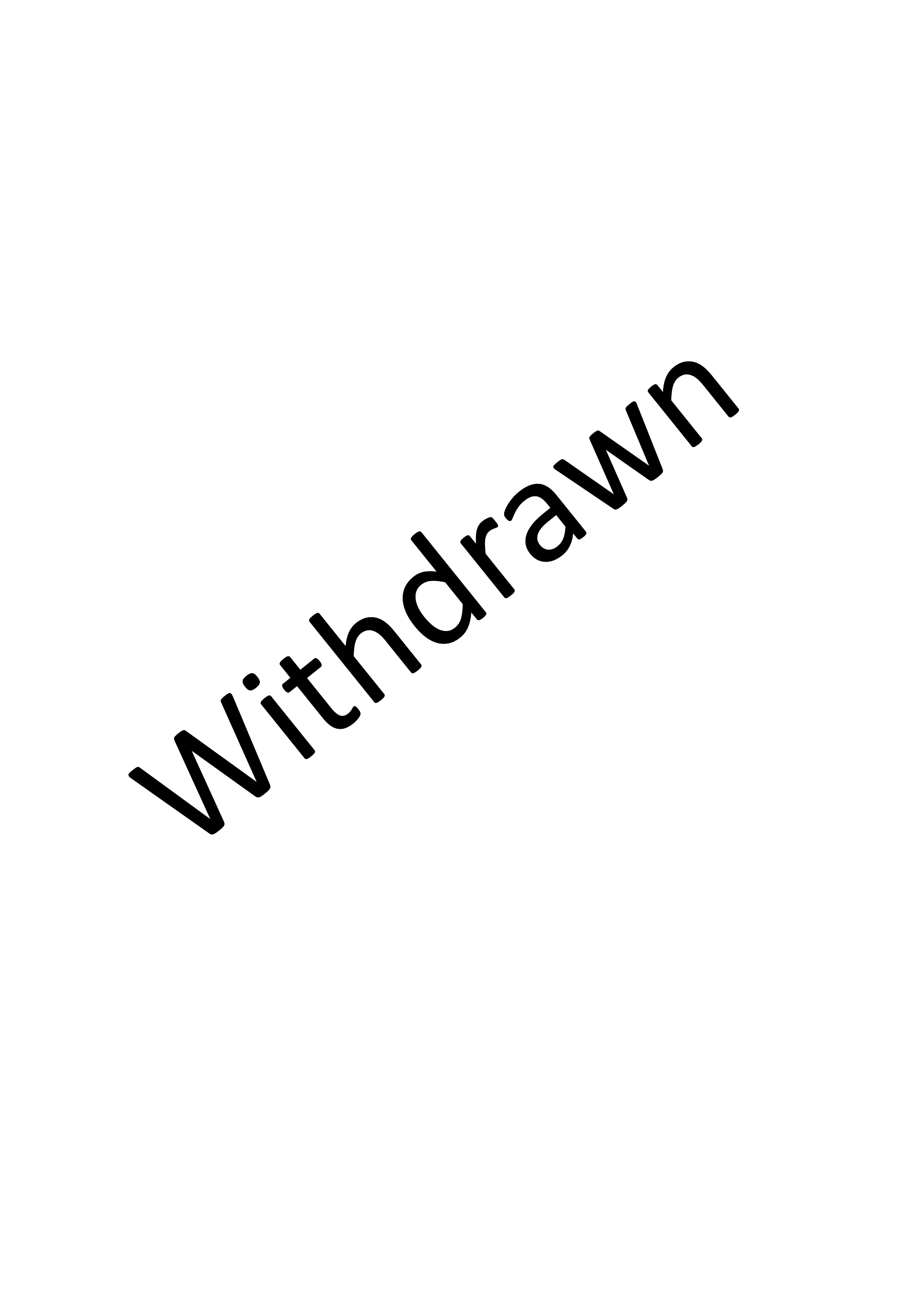}}}
\def\checkmark{\tikz\fill[scale=0.4](0,.35) -- (.25,0) -- (1,.7) -- (.25,.15) -- cycle;} 

\ifCLASSINFOpdf

\else

\fi
\newcommand{\ds}{\displaystyle}

\begin{document}

\title{Analyzing Insect-Plant Predation Data By Bayesian Nonparametrics}

\author{\IEEEauthorblockN{Fan Yang, Takatomi Kubo, and Kazushi Ikeda}}

\maketitle

\begin{abstract}

In the prospect of ecology and biology, studying insect-plant predation will considerably contribute to pest control, benefit agriculture and afforestation, and also help people to better understand insect-plant co-evolution. Therefore, we are motivated to do two work in this study. The first part is to cluster the insect-plant predation, in such manner, unobserved predation could be estimated. The second part is to explore the connection between predation and bio-taxonomy, and we find insects get more divergence than plants during the insect-plant co-evolution (sorry for withdraw it). \\

\end{abstract} 

\IEEEpeerreviewmaketitle

\section{Introduction}

Predation is a relationship that a predator eat its prey, and it is variant from species to species. In the case of herbivorous insects and plants, a species of insect eats one or a few specific species rather than any plant. This predation pattern has been developed during their co-evolution \cite{selection,eatbehavior,evolution1,evolution2}. Due to the immense number of insect species and plant species, the size of predation is incredibly huge. However, since the predation data is observed by biologists one by one, it poses a problem: only a very limited number of observed predation data is available, it is hard to study each individual independently.
To tackle this problem, we cluster the predation data, in such way, common properties are enriched by grouping similar individuals together, which can help us in further analysis.
Furthermore, because the number of clusters in insects and plants is unknown,
we applied a Bayesian nonparametric \cite{nonpara} to the analysis
as we applied to web data \cite{konishi} or driving behaviors \cite{tiv}.


\section{Data}

\subsection{Insect-plant predation data}

Our observed data consist of 615 insect species and 1273 plant species,
that is 615 by 1273 pairs of insect-plant predation in total.
To represent this relationship, we used a bipartite graph,
where an edge between two nodes, one for an insect species and the other for a plant species,
expressed a predation observation.
Note that no edge does not mean non-predation but just non-observation,
in other words, this graph had missing edges.

The bipartite graph can be expressed as a matrix
where the value one meant the existence of an edge and the value zero meant no edge
(TABLE \ref{relation}).
Our data had 2478 ones in 615 by 1273 components.
\begin{table}[h]
\captionsetup[table]{skip=10pt}
\centering 
\begin{tabular}{|c|c|c|c|c|}
\hline
         & Plant 1 & Plant 2 & Plant 3 & ... \\ \hline
Insect 1 & 0       & 0       & 1       &     \\ \hline
Insect 2 & 1       & 0       & 0       &     \\ \hline
Insect 3 & 0       & 0       & 1       &     \\ \hline
...      &         &         &         &     \\ \hline 
\end{tabular}
\caption{Insect-plant predatory relationship representation}
\label{relation}
\end{table}

\subsection{Bio-taxonomic data}
Bio-taxonomy is a classification that divides creatures into hierarchical groups based on biological features (Fig.~1) \cite{taxonomy}.
To some extent, this descending order coincides with life evolutionary history. 
For example, in bio-taxonomy, human and apes are belong to different species but the same family (i.e., Hominidae).
This reflects the fact that they shared the same ancestor millions of years ago.
\begin{figure}[!h]
\centering
\includegraphics[width=46mm]{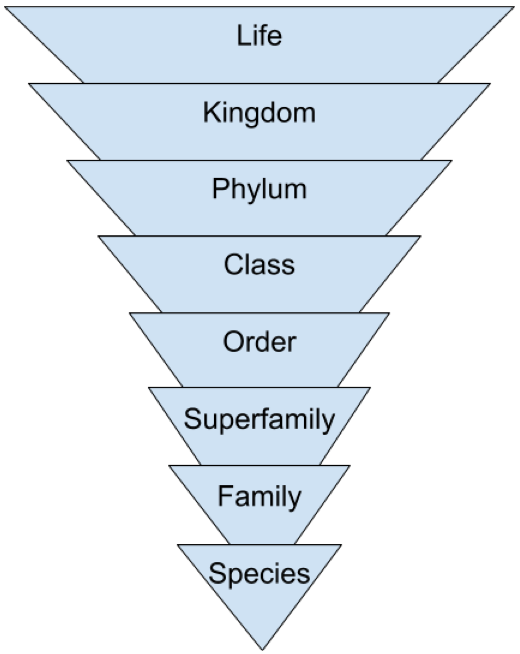}
\caption{\label{fig:taxonomy} Descending order of bio-taxonomy}
\end{figure}

In our analysis, we employed the bio-taxonomic information from \cite{},
to 615 insect species and 1273 plant species.
The available bio-taxonomic information for our data is shown in TABLE \ref{taxonomy}.
\begin{table}[!h]
\captionsetup[table]{skip=10pt}
\centering
\begin{tabular}{|c|c|c|}
\cline{2-3}
\hline                                   			& Insect 	& Plant \\ \hline
\multicolumn{1}{|c|}{Kingdom}      	&        		& \checkmark      \\ \hline
\multicolumn{1}{|c|}{Class}        		&  \checkmark      &       \\ \hline
\multicolumn{1}{|c|}{Order}        		&  \checkmark      & \checkmark   \\ \hline
\multicolumn{1}{|c|}{Superfamily}       &  \checkmark      & 		   \\ \hline
\multicolumn{1}{|c|}{Family}       	&   \checkmark   &   \checkmark    \\ \hline
\multicolumn{1}{|c|}{Species}      	&  \checkmark    &  \checkmark      \\ \hline
\end{tabular}
\caption{Bio-taxonomies available in this work}
\label{taxonomy}
\end{table}
\\

\section{Mathematical model for coclustering}

\subsection{Infinite relational model}

The Infinite Relational Model (IRM) is a Bayesian nonparametric model for clustering relational data \cite{IRM}.
By using Chinese Restaurant Process (CRP) as a prior,
the number of clusters is adjusted by the data structure automatically \cite{CRP}.
In our case, cluster assignments for the first domain (insects) $X=\{x_{1}, \ldots, x_{n}\}$
and the second domain (plants) $Y=\{y_{1}, \ldots, y_{m}\}$, are generated by two CRP separately,
that is,
\begin{align}
  z^{x}_{i} &\sim \mathrm{CRP}(\alpha_{1}), \quad i=1, \ldots, n, \\
  z^{y}_{j} &\sim \mathrm{CRP}(\alpha_{2}), \quad j=1, \ldots, m,
\end{align}
where $\alpha_{1}$ and $\alpha_{2}$ are the concentrated hyper parameters in CRP\null.

The edge between two clusters, $k$ and $l$, is generated
according to the Bernoulli distribution with probability $\eta_{kl}$.
While $\eta=\{\eta_{kl}\}_{k=1,l=1}^{\infty,\infty}$, is generated
from the Beta distribution with parameters, $a$ and $b$,
that is,
\begin{align}
  \eta  &\sim \mathrm{Beta}(a,b), \\ 
  R_{ij} &\sim \mathrm{Bernoulli}(\eta_{z^{x}_{i}z^{y}_{j}}),
\end{align}  
where $R_{ij}$ denotes a binary variable representing the existing of the edge between nodes, $i$ and $j$.

\subsection{Bayesian inference for IRM}

In order to co-clustering the insect-plant predatory relational data,
we need to estimate the assignments, $z^x_i$ and $z^y_j$,
as well as the link probabilities, $\eta=\{\eta_{kl}\}$, from the observation, $R_{ij}$.

Since our generative model has a conjugate prior distribution,
we can integrate $\eta$ out analytically.
In fact, $\eta$ is updated as
\begin{align}
\eta_{kl}=\frac{a+n(kl)}{a+b+n(kl)+\bar{n}(kl)},
\end{align}
where $n(kl)$ is the number of the node pairs between $k$ and $l$ that have a link,
$\bar{n}(kl)$ is the number of those that do not have a link.

In contrast, we need a sampling method to calculate the cluster assignments.
In our experiments below, the Collapsed Gibbs Sampling \cite{IRM_model} is employed,
the probability for the first domain is expressed as
\begin{align}
  & P(z^{x}_{i} = k | z^{X}_{-i}, z^{Y}, R; \alpha_{1}, \alpha_{2}, a, b)\propto \nonumber\\
  &
  \begin{cases}
    N_{k,-i} \prod_{l} \frac{\ds B(a+n_{+i}(kl), b+ \bar{n}_{+i}(kl))}%
    {\ds B(a+n_{-i}(kl), b+ \bar{n}_{-i}(kl))} 
    & k \leq K, \\  
    \alpha_{1} \prod_{l} \frac{\ds B(a+n_{+i}(kl), b+ \bar{n}_{+i}(kl))}{\ds B(a,b)}
    & k = K+1,
  \end{cases}
\end{align}
where $N_{k}$ denotes the number of objects in cluster $k$,
and $K$ is the number of existing clusters.
Note in all variables, the $-i$ means excluding object $x_{i}$,
while the $+i$ means adding object $x_{i}$.

In the same way, the probability for the second domain is expressed as
\begin{align}
  & P(z^{y}_{j} = l | z^{Y}_{-j}, z^{X}, R; \alpha_{1}, \alpha_{2}, a, b)\propto \nonumber\\
  &
  \begin{cases}
    N_{l,-j} \prod_{k} \frac{\ds B(a+n_{+j}(kl), b+ \bar{n}_{+j}(kl))}%
    {\ds B(a+n_{-j}(kl), b+ \bar{n}_{-j}(kl))}
    & l \leq L \\  
    \alpha_{2} \prod_{k} \frac{\ds B(a+n_{+j}(kl), b+ \bar{n}_{+j}(kl))}{\ds B(a, b)}
    & l = L+1,
  \end{cases}
\end{align}
where $N_{l}$ denotes the number of objects assigned to cluster $l$
and $L$ is the number of existing clusters. The $-j$ means excluding object $y_{j}$,
while the $+j$ means adding object $y_{j}$.\\

\section{Experiments}

\subsection{Model setting}

We co-clustered our insect-plant predation data
using the Bayesian inference of IRM in the previous section.
Since the Collapsed Gibbs Sampling is one of MCMC methods, it assures asymptotic convergence to the true posteriors when the number of samples going to infinite \cite{convergence}, and diagnosing convergence is difficult. In real practice, infinite iteration is impracticable, so we make a trade-off between computing cost and clustering quality. In our study, we use 300 iterations, and set the first 200 iterations as burn-in data, namely, to be abandoned.

\subsection{Results}

We estimated the assignments and the probabilities 
by a public Matlab code \cite{code}.
As mentioned before, the inference was stochastic in sampling, so the clustering results were not uniquely determined. Despite that, over several runs, we consistently find the number of insect-clusters ranges from 9 to 10, while the number of plant-clusters ranges from 7 to 8. In further discussion, we show this kind of clustering variety does not affect the analysis that we focus on.

\begin{figure*}[!tbp]
  \centering
  \begin{subfigure}[b]{0.48\textwidth}
    \includegraphics[width=\textwidth]{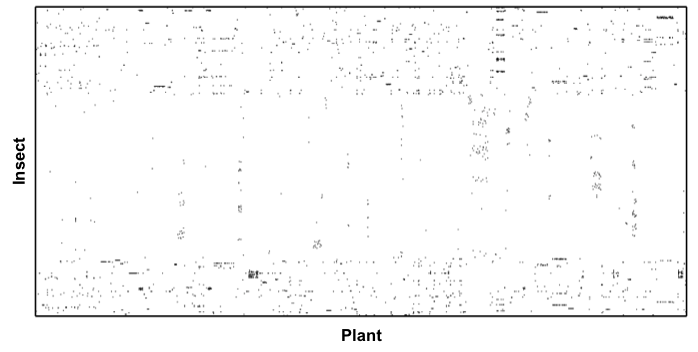}
    \caption{Raw Predation Data}
  \end{subfigure}
  \hfill
  \begin{subfigure}[b]{0.48\textwidth}
    \includegraphics[width=\textwidth]{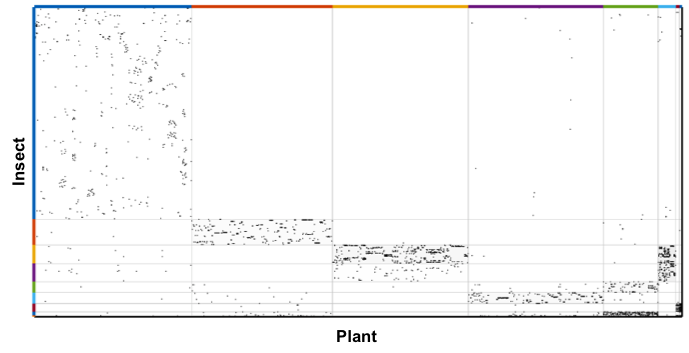}
    \caption{Sorted Predation Data According To IRM Cluster Assignments}
  \end{subfigure}
  \caption{\label{CoclusteringResults}
  An example of co-clustered insect-plant predation}
\end{figure*}

\begin{figure*}[h!]
\centering
\includegraphics[width=65mm]{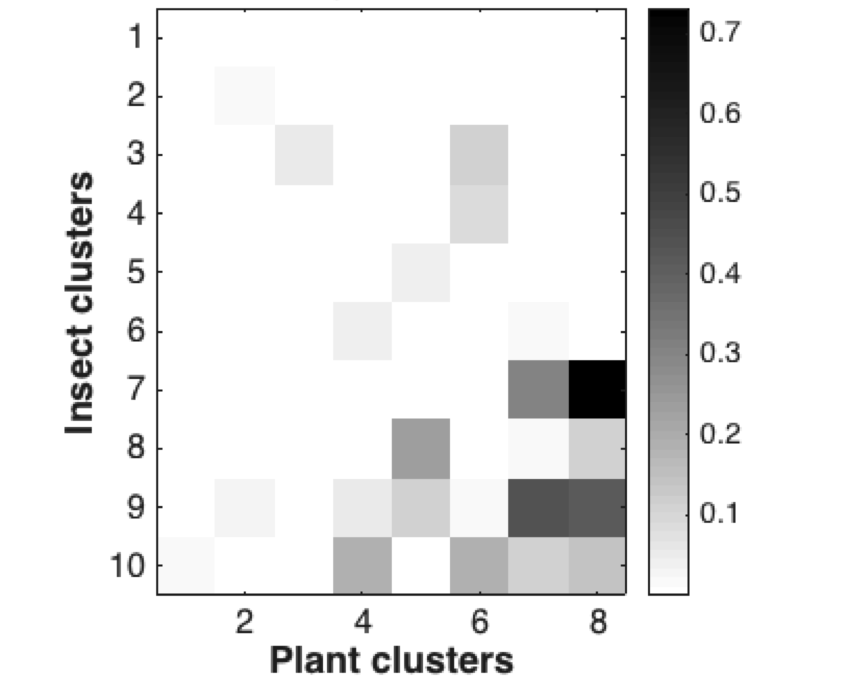}
\caption{\label{etaMatrix}
  The $\eta$ matrix expresses the densities of insect-plant cluster pairs.}
\end{figure*}

\begin{figure*}[h!]
\centering
\includegraphics[width=120mm]{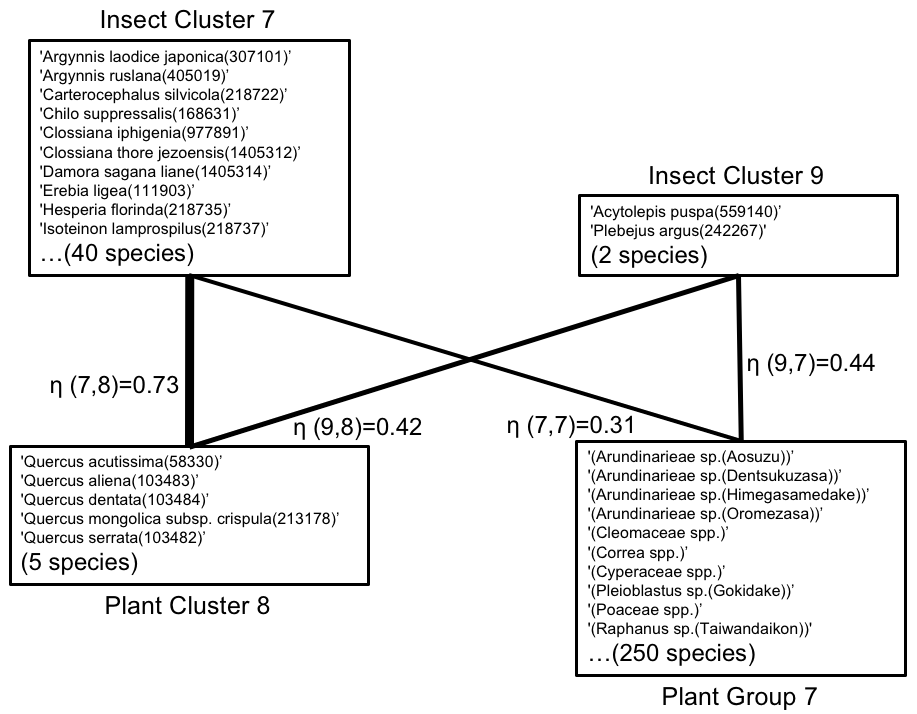}
\caption{\label{fig:top4} Cluster pairs of top 4 maximum $\eta$ value}
\end{figure*}

The illustration in Fig.~\ref{CoclusteringResults} is one of the various results when insects
are split to 10 clusters, while plants are split to 8 clusters.
Entities that belong to the same cluster are sorted together, so we can see the density of
observed data is higher in specific insect-plant cluster pair. 
A quantitative representation of these density difference is the $\eta$ matrix (Fig.~\ref{etaMatrix}).
To estimate which insects  are likely to eat which plants
even though they are not observed, we can refer to the cluster assignment and $\eta$ matrix.
For example, if we choose the top four maximum $\eta$ values among the cluster pairs,
we can see the member of each clusters and the interactions within them (Fig.~\ref{fig:top4}).

\begin{figure*}[h!]
\centering
\includegraphics[width=120mm]{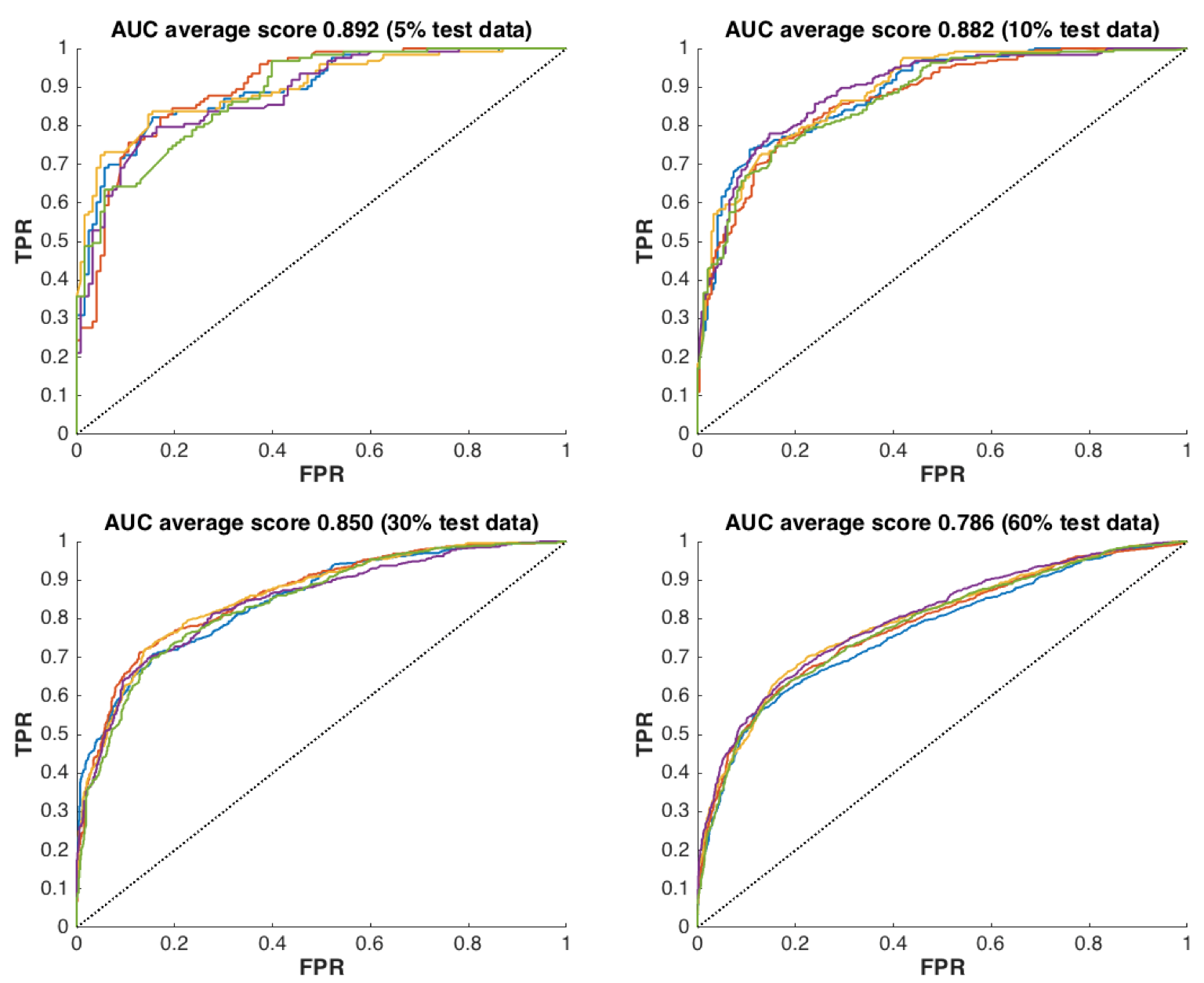}
\caption{\label{fig:AUC} AUC of ROC testing results.}
\end{figure*} 

\begin{figure*}[h!]
\centering
\includegraphics[width=140mm]{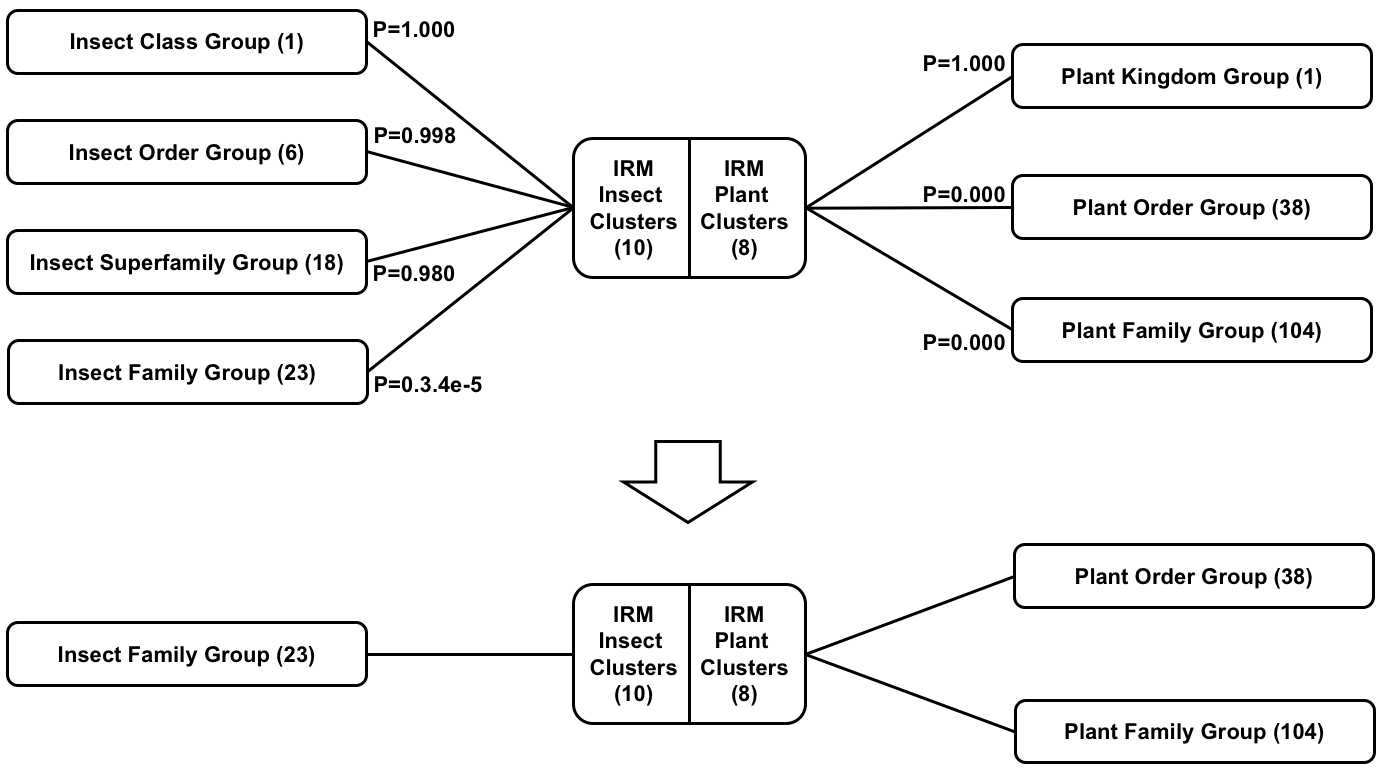}
\caption{\label{fig:chi_test} The significance of association test ($p<0.05$).}
\end{figure*}

\subsection{Evaluation}

Because the ground truth of insect-plant predation clusters does not exist, 
it may be hard to convince ourselves that IRM generates a trustworthy result without evaluation. 
To allay such worries, we suppose the clustering quality is identical to the quality of predicting missing relationship. 
In such manner, we alternatively assess the quality of predicting missing relationship by 
cluster assignment and $\eta$ value.

We randomly choose a proportion of data with value $1$, coupled with the same amount of  data with value $0$, to be the testing set, which  will not participate in IRM process. 
After IRM generates clusters, we evaluate how well the entities with original label `1' and `0' in the testing set can be separated, according to the cluster assignment and $\eta$. The performance will be measured by
the receiver operator characteristic (ROC). Through calculating the area under curve (AUC) of ROC,  the AUC score will range from 0.5 to 1.0, indicating the performance of estimation from poor to perfect, respectively.

Choosing testing sets proportion as 5\%, 10\%, 30\% and 60\%, we get AUC score as
0.892, 0.882, 0.850 and 0.786, respectively (Fig.\ref{fig:AUC}). 
Note each score is the average performance
of five times tests, by randomly splitting data set into the training set and testing set.
Hence, we conclude, to some extent, IRM has correctly clustered our data.

\section{Discussion}

Predatory behaviours of animals are considered to correlate with bio-taxonomy under family level \cite{prey_taxonomy},
because under family level, animal species within the same group usually share many common biological properties.
Nonetheless, this conclusion was based on an intuitive analysis.
In addition, only the predator in the predation was considered
and the other element, the prey such as plants, was not taken into account.

Since our analysis co-clustered both the predator and the prey,
it is possible to analyze the dependency between the bio-taxonomies in both groups.
Thus, we applied the $\chi^2$-test ($p<0.05$) to their correlation
to see if they were correlated or not (Fig.~\ref{fig:chi_test}) \cite{{chi_squared}}.

For insects, the $p$-value drastically increased from Insect Family to Insect Superfamily.
This means that predation correlates with the bio-taxonomy only at family level,
which is consistent with the existing conclusion \cite{prey_taxonomy}.

Similarly, we found the correspondence between Plant Family groups and IRM Plant clusters,
Plant Order groups and IRM Plant clusters.
This means that plants share more similarity even at a high level of the bio-taxonomy
compared with insects. Note the result of this statistical test held irrespective of the repeated experiments.
Although the IRM clustering is stochastic, $\chi^2$-test results are consistent.\\

\section{Conclusions} 

In this paper, we demonstrated that the co-clustering method based on the IRM is useful
to analyze predation data, even when only a small part of data are available.
And we can use this result to estimate unobserved predation.
Besides, based on the clustering result, we also found that
insects only correlate with bio-taxonomy at family level,
while plants have dependency at many levels.
This knowledge suggests that
insects get more divergence than plants during the insect-plant co-evolution.


\begin{thebibliography}{1}

\bibitem{selection}
  Waldbauer, G. P., Friedman, S.:
  Self-selection of optimal diets by insects,
  \textit{Annual Review of Entomology}, \textbf{36}/1 (1991), pp. 43--63.
 
\bibitem{eatbehavior}
  Schoonhoven, L.M.:
  What makes a caterpillar eat? The sensory code underlying feeding behavior,
  In \textit{Perspectives in Chemoreception and Behavior}, Springer, New York, 1987, pp. 69--97.

\bibitem{evolution1}
  Whitney, H.M., Glover, B.J.:
  Coevolution: Plant-insect,
  In \textit{Encyclopedia of Life Sciences},  Wiley Online Library, 2013.

\bibitem{evolution2}
  Jermy, T.:
  Evolution of insect/host plant relationships,
  \textit{American Naturalist}, \textbf{124}/5 (1984), pp. 609--630.

\bibitem{nonpara}
  Hjort, N.L., et al., ed.:
  \textit{Bayesian Nonparametrics}, Cambridge Series in Statistical and Probabilistic Mathematics, 2010.

\bibitem{konishi}
  Konishi, T., et al.:
  Variational Bayesian inference algorithms for infinite relational model of network data,
  \textit{IEEE Trans. Neural Networks and Learning Systems}, \textbf{26}/9 (2015), pp. 2176--2181.

\bibitem{tiv}
  Hamada, R., et al.:
  Modeling and prediction of driving behaviors based on automatic temporal segmentation,
  \textit{IEEE Trans. Intelligent Vehicles}, in press.

\bibitem{taxonomy}
  Mayr, E.:
  \textit{The Growth of Biological Thought: Diversity, Evolution, and Inheritance}. Harvard University Press, 1982.

\bibitem{}
  A new reference.

\bibitem{IRM}
  Kemp, C., et al.:
  Learning systems of concepts with an infinite relational model,
  In \textit{Proc. AAAI}, pp. 381--388, 2006.

\bibitem{CRP}
  Murphy, K.P.:
  \textit{Machine Learning: A Probabilistic Perspective}, MIT press, 2012. 


\bibitem{IRM_model}
  Ishiguro, K., et al.:
  Collapsed variational Bayes inference of infinite relational model,
  arXiv:1409.4757, 2014.
  
  \bibitem{code}
  Morten M{\o}rup, Exercise material on the infinite relational model,
  Available at \texttt{http://www.mortenmorup.dk/index\_files/}
 \texttt{Page327.htm},
  accessed on Nov. 4, 2016.

\bibitem{convergence}
  Cowles, M.K., Bradley P.C.:
  Markov chain Monte Carlo convergence diagnostics: a comparative review,
  \textit{J. American Statistical Association}, \textbf{91}/434 (1996), pp. 883--904.

\bibitem{prey_taxonomy}
  Boucot, A.J.:
  \emph{Evolutionary Paleobiology of Behavior and Coevolution}, Elsevier, 2013.

\bibitem{chi_squared}
  Greenwood, P.E., Nikulin, M.S.:
  \emph{A Guide to Chi-Squared Testing}, John Wiley \& Sons, 1996.

\end{thebibliography}
\end{document}